%% file: uncertainty_financial_shocks.tex
\documentclass[notitlepage]{article}
\usepackage{arxiv}
\usepackage[utf8]{inputenc}
\usepackage{lastpage}   
\usepackage{amsmath}
\usepackage{amssymb}
\usepackage[english]{babel}
\usepackage{import}

\usepackage{mathtools}
\usepackage{amsfonts}
\usepackage{amssymb}
\usepackage{caption}
\usepackage{subcaption}
\usepackage{appendix}
\usepackage{graphicx}
\usepackage[authoryear, sectionbib]{natbib}
\usepackage{titling}
\usepackage{chapterbib}  
\usepackage[nameinlink, capitalize]{cleveref} 
\usepackage{etoolbox} 
\usepackage[flushleft]{threeparttable}
\usepackage{tabularx}
\usepackage{multirow}
\usepackage{booktabs,caption,fixltx2e}
\usepackage{epstopdf}
\usepackage{rotating}
\usepackage{url}
\crefname{appsec}{Appendix}{Appendix}
\usepackage{float}

\addto{\captionsenglish}{%
  
}%

\newcommand{\matr}[1]{\mathbf{#1}}

\DeclarePairedDelimiterX{\norm}[1]{\lVert}{\rVert}{#1}

\makeatletter

\patchcmd{\NAT@test}{\else \NAT@nm}{\else \NAT@nmfmt{\NAT@nm}}{}{}

\DeclareRobustCommand\citepos
  {\begingroup
   \let\NAT@nmfmt\NAT@posfmt
   \NAT@swafalse\let\NAT@ctype\z@\NAT@partrue
   \@ifstar{\NAT@fulltrue\NAT@citetp}{\NAT@fullfalse\NAT@citetp}}

\let\NAT@orig@nmfmt\NAT@nmfmt
\def\NAT@posfmt#1{\NAT@orig@nmfmt{#1's}}

\makeatother
\makeatletter
\def\@makechapterhead#1{%
  {\parindent \z@ \raggedright \normalfont
    \ifnum \c@secnumdepth >\m@ne
        \huge\bfseries \thechapter\ 
    \fi
    \hskip 20\p@
    \interlinepenalty\@M
    #1\par\nobreak
    \vskip 40\p@
  }}
  
 \def\@makeschapterhead#1{%
  \vspace*{-50\p@}
  {\parindent \z@ \raggedright
    \normalfont
    \interlinepenalty\@M
    \Huge \bfseries  #1\par\nobreak
    \vskip 40\p@
  }}
  
\makeatother

\providecommand{\keyword}[1]
{
  \textbf{\textit{Keywords ---}} #1
}

\providecommand{\jelcodes}[1]
{
  \textbf{\textit{JEL Codes ---}} #1
}

\numberwithin{equation}{section}    
\numberwithin{table}{section}       
\numberwithin{figure}{section}      

\shorttitle{Macroeconomic Effects of Uncertainty and Financial Shocks}

\title{Macroeconomic Effect of Uncertainty and Financial Shocks: a non-Gaussian VAR approach\thanks{I gratefully acknowledge financial support from the Academy of Finland (grant 308628)}.}

\author{Olli Palmén\thanks{E-mail: olli.palmen@helsinki.fi}\\
	University of Helsinki\\
}

\renewcommand{\headeright}{Working Paper}
\renewcommand{\undertitle}{Working Paper}

\usepackage{titling}

\pretitle{\begin{center}\Large\bfseries}
\posttitle{\par\end{center}\vskip 0.5em}
\preauthor{\begin{center}\large}
\postauthor{\end{center}}
\predate{\par\large\centering}
\postdate{\par}

\date{\today}

\begin{document}

\maketitle

\begin{abstract}
\input{./tex/front/abstract}
\end{abstract} \hspace{20pt}

\keyword{financial shocks, economic uncertainty, structural vector autoregression, business cycle fluctuations} \hspace{10pt}

\jelcodes{E23, E31, E32, E44, E52}

\graphicspath{{./tex/ch3/img/}}
\include{./tex/ch3/ch3}


\cleardoublepage %

\end{document}

%% file: tex/front/abstract.tex
The Great Recession highlighted the role of financial and uncertainty shocks as drivers of business cycle fluctuations. However, the fact that uncertainty shocks may affect economic activity by tightening financial conditions makes empirically distinguishing these shocks difficult. This paper examines the macroeconomic effects of the financial and uncertainty shocks in the United States in an SVAR model that exploits the non-normalities of the time series to identify the uncertainty and the financial shock. The results show that macroeconomic uncertainty and financial shocks seem to affect business cycles independently as well as through dynamic interaction. Uncertainty shocks appear to tighten financial conditions, whereas there appears to be no causal relationship between financial conditions and uncertainty. Moreover, the results suggest that uncertainty shocks may have persistent effects on output and investment that last beyond the business cycle.

%% file: tex/ch3/ch3.tex
\section{Introduction} \label{ch3:sec:introduction}
               
Following the Great Recession of 2007--2009, economists have come to view uncertainty and financial shocks as important sources of business cycle fluctuations \citep{Christiano2014, Bloom2018, Arellano2018}. A challenge in empirically assessing the macroeconomic effect of uncertainty and financial shocks is that they are difficult to distinguish in practice. Uncertainty shocks---generally defined as unexpected changes in the conditional volatility of economic variables observed by economic agents---affect economic activity by reducing consumption due to precautionary saving (\cite{Leland1978,Kimball1990}), postponing investment decisions (\cite{Bernanke1983} or tightening financial constraints in the presence of financial frictions (\cite{Gilchrist2014}).

Given that uncertainty shocks affect economic activity in part by tightening financial constraints, both the financial shock and the uncertainty shock are associated with higher credit spreads and a decrease in credit supply. Specifically, strong correlation between measures of uncertainty and financial risk and the fact that both shocks affect economic and financial variables simultaneously makes it difficult to employ standard methods, such as recursive ordering or sign restrictions, to identify the shocks in a structural vector autoregression (SVAR) framework.

Recent studies examine the effects of uncertainty and financial shocks on the business cycle within the same framework. \cite{Stock2012} use instrumental variables to identify the financial and the uncertainty shock in a factor-augmented vector autoregressive (FAVAR) framework, although they find that strong correlation between instrumental variables for the financial shocks and the uncertainty shocks prohibits the distinct identification of these shocks. \cite{Caldara2016} identify the financial and uncertainty shocks in a SVAR model by using a penalty function approach, following \cite{Faust1998}, \cite{Uhlig2005} and \cite{Mountford2009}. This approach requires that shocks are identified in a given sequence, making the identification conditional on the order in which the shock is identified. Hence, this approach is only able to provide bounds for the macroeconomic effects of these shocks. In a smilar setting, \cite{Furlanetto2019} simultaneously identify financial and uncertainty shocks along with other structural shocks using sign restrictions.

Although statistical identification been used to study the macroeconomic effcts of the financial shock (\cite{Brunnermeier2020}) and the uncertainty shock (\cite{Carriero2021}) separately, these types of identification techniques have not been previosly used to identify uncertainty and financial shocks simultaneously. To this end, we propose a new approach to identifying the financial and uncertainty shock based on non-Gaussianity by \cite{Anttonen2021}, which builds on the pioneering work \cite{Lanne2017}. The structural shocks are first identified based on their distributional properties, and the structural shocks are given economic labels based on the properties of the impulse response functions. That is, we label the uncertainty and financial shocks to be the shocks whose response on impact to uncertainty and financial conditions is greatest on impact, respectively\footnote{It is possible that either none or two or more shocks satisfy these conditions, in which case other extraneous information needs to be used to label the structural shocks.}.


Labelling the shocks based on the properties of the impulse responses is conceptually similar to the penalty-function approach, in which structural shocks are identified by by minimizing a predetermined penalty function. Our identification approach is therefore similar to that of \cite{Caldara2016}, but it has the advantage that the structural shocks are uniquely identified and it does not require restrictions \textit{a priori}.

We study the macroeconomic effects of the financial and uncertainty shocks in the United States using both monthly and quarterly data. Although the financial and uncertainty shocks are commonly estimated using monthly data, we estimate the model using quarterly data to capture their long run effects and to assess the robustness of the estimates obtained using a higher frequency of observations. The results imply that macroeconomic uncertainty and financial shocks are both independent sources of business cycle fluctuations that also interact dynamically. Namely, the uncertainty shock seems to affect economic activity at least partly by tightening financial constraints, and the financial shock may also be amplified by an increase in uncertainty. Moreover, the results imply that the uncertainty shock may affect the economy beyond the business cycle by inducing a persistent negative effect on output and investment by even more than documented in the previous literature (e.g., \cite{Bonciani2019, Furlanetto2019}).

The rest of the paper is organized as follows. \Cref{ch3:sec:literature} gives an overview of related literature. \Cref{ch3:sec:methods} outlines the econometric method. \Cref{ch3:sec:macroeffects} describes the macroeconomic effects of uncertainty and financial shocks. \Cref{ch3:sec:conclusion} summarizes the results and concludes.

\section{Empirical literature on financial and uncertainty shocks}  \label{ch3:sec:literature}

The literature on macroeconomic effects of financial and uncertainty shocks has grown considerably in recent years. Much of the existing empirical literature have studied the effects of uncertainty and financial shocks on economic activity separately, but only few efforts have been made to study the interaction between financial conditions and uncertainty.

Empirical studies have used information contained in credit spreads of financial contracts to estimate the business cycle effects of financial shocks in the SVAR framework. However, these studies do not explicitly take into account the effect of macroeconomic uncertainty, which is strongly correlated with financial conditions. \cite{Gilchrist2009} identify the financial shock by using corporate credit spreads in a FAVAR framework. \cite{Gilchrist2012} construct a measure of cyclical changes in default risk and credit spreads, the excess bond premium, and use this measure to study the macroeconomic effects of the financial shock in a recursively identified vector autoregressive (VAR) model. \cite{Brunnermeier2020} study the macroeconomic effects of financial shocks in a structural VAR model by using heteroskedasticity and non-normality for identification. They find that the credit spread shock originating in the credit spreads and the credit spread shock originating in the interbank lending spreads both have strong negative effects on real economic activity.


A number of empirical studies also assess the macroeconomic effects of uncertainty shocks in the SVAR framework, but they often do so in isolation of financial effcts. These studies typically employ different measures of uncertainty, but are often identified using the recursive ordering approach (e.g., \cite {Bachmann2013, Baker2016, Bonciani2019, Caggiano2014, Caldara2018, Jurado2015, Leduc2016, Rossi2015}). However, these studies provide mixed evidence of the macroeconomic effects of uncertainty shocks and the direction of causality between uncertainty and economic activity. Inconsistent findings may be due to the sensitivity of the results to the identifying assumptions and the problems with addressing simultaneous relationships between macroeconomic variables in recursive models \citep{Baker2013}. Moreover, \cite{Carriero2015} find that including a measure of uncertainty as an endogenous variable may leads to biased impulse response functions, if the proxy variable includes a measurement error.

Recent studies have attempted to overcome the problems with recursive identification schemes by using measures of uncertainty as external instruments in the proxy-SVAR framework (e.g., \cite{Baker2016, Carriero2015, Stock2012, Piffer2018}). For example, \cite{Carriero2015} find larger and more persistent negative effects of the uncertainty shock when the measure of uncertainty is included as an external instrument compared to recursively identified models. Other recent studies use sign restrictions and narrative restrictions to identify the uncertainty shock, while preserving the assumption that uncertainty may be subject to feedback from macroeconomic variables. \citeauthor{Ludvigson}(forthcoming) include measures of macro and financial uncertainty as endogenous variables alongside real economic activity. They identify macroeconomic and financial uncertainty shocks using "shock-based restrictions", where the set of admissible shocks is obtained from a large number of orthonormal shocks based on criteria related to the characteristics of the shocks during historical episodes. They find the financial uncertainty shock to be associated with persistent negative effects on real activity. However, their results suggest that the macroeconomic uncertainty shock does not cause business cycle fluctuations, but may likely amplify them. \cite{Shin2020} study the macroeconomic effects of macroeconomic uncertainty and financial uncertainty shocks in SVAR with stochastic volatility using sign restrictions. \cite{Carriero2021} exploit time-varying volatility in the time 	series in SVAR model with stochastic volatility to identify uncertainty shocks in the SVAR framework, following the results by \cite{Lewis2021}. They find evidence of feedback effects between uncertainty and macroeconomic variables, suggesting that uncertainty is endogenous.

Several studies have also considered the interplay between financial conditions and uncertainty as sources of business cycle fluctuations, following the theoretical contributions by \cite{Christiano2014} and \cite{Arellano2018} among others. These studies generally find that both the uncertainty shock and the financial shock have a negative effect on the economic activity, with uncertainty shocks affecting real economic activity through a tightening in financial conditions. We contribute to this work by identifying the structural shocks based on statistical identification techniques.

In a closely related paper, \cite{Caldara2016} identify the financial and uncertainty shocks in an SVAR model, using the penalty-function approach as pioneered by \cite{Uhlig2005} and monthly data for the US. They define the penalty-function such that the uncertainty shock and the financial shock maximize the impulse response of uncertainty and financial conditions over a given horizon, respectively. In a similar setting, \cite{Dery2021} study the relative importance of financial, uncertainty and monetary policy shocks in the US in the SVAR framework. Their model includes measures for financial conditions, uncertainty and monetary policy. They identify the structural shocks based on sign restrictions using the penalty-function approach, where each structural shock is restricted to have a positive effect on the target variable for a given horizon following the shock and no effect on the other target variables. In a related study \cite{Furlanetto2019} use sign restrictions in the SVAR framework to identify financial, uncertainty and housing-market shocks using quarterly data for the US. They disentangle the financial and uncertainty shocks by restricting the response of the adjusted ratio of an uncertainty measure to financial conditions, such that a positive response is associated with an uncertainty shock and a negative response is related to a financial shock. The scope of our work bares similarities with these studies, but the novel identification approach has the advantage that the structural shocks are uniquely identified and it does not require strong \textit{a priori} restrictions. Moreover, it extends the previous analysis of \cite{Caldara2016} by estimating the model using quarterly US data. 

\cite{Popp2016} study the direct and indirect macroeconomic effects of the uncertainty shock and the relative importance of the financial channel in different economic regimes. They estimate a smooth-transition factor-augmented vector-autoregression (ST-FAVAR) model using monthly data for the US and decompose the total effect of the uncertainty shock into a direct and an indirect effect affecting credit conditions using a counter-factual decomposition similar to \cite{Sims2006}. They find uncertainty shocks to have short-run negative effects on both real economic activity and stock returns. Moreover, they report a greater negative effect on economic activity during recessions, when the importance of the financial channel is also greater. This is consistent with the findings of \cite{Alessandri2019} who identify the uncertainty shock as an exogenous innovation in the volatility of structural shocks in different financial regimes. 


\section{Bayesian estimation of non-Gaussian SVARs}  \label{ch3:sec:methods}


SVAR models conventionally assume that the shocks follow a Gaussian distribution. In a Gaussian setting, restrictions based on economic theory help to identify the structural shocks of interest by narrowing down the set of structural shocks. For example shocks may be identified by imposing restrictions on the signs of the response of variables. However, recently new methods have been introduced to uniquely identify the structural shocks in the case that the error process is non-Gaussian. These methods are applicable to macroeconomic time series, which often exhibit skewness or heavy tails. Moreover, the identification of the structural shocks is based on the statistical properties of the time series and therefore no additional restrictions on the relationship between variables or the data need not be imposed. However, since statistically identified shocks do not by themselves have an economic interpretation, additional information is needed to label the shocks. For example, checking that the sign of the response of variables to a given shock corresponds to theoretical predictions may be sufficient to give in an economically meaningful label.

We estimate our model using a Bayesian framework for non-Gaussian SVARs proposed by \cite{Anttonen2021}. The estimation is based on the identification method proposed by \cite{Lanne2017} and \cite{Lanne2020} by introducing a more general distributional assumptions for the shocks, namely that the shocks follow a skewed generalized t distribution (henceforth, sgt-distribution). This reduces the risk of misspecification and potentially strengthens the identification of the model.

\subsection{The SVAR model with potentially skewed and fat-tailed errors}

We consider the general structural vector autoregression model

\begin{equation}
y_t = c + A_1 y_{t-1} + \cdots + A_p y{t-p} + B \epsilon_t ,
\label{ch3:eq:SVAR}
\end{equation}

\noindent where $y_t$ is the $1 \times n$ vector of observations at time $t$, $c$ is the intercept, $A_1, \ldots, A_p$ are the $n \times n$ coefficient matrices  and $B$ is the  $n \times n$ non-singular matrix that describes the contemporaneous impact of the structural shocks $\epsilon_t$. 

For the purpose of identifying the structural shocks, we assume that each component $\epsilon_{it}$ of $\epsilon_t= (\epsilon_{1t}, \ldots, \epsilon_{nt})'$ follows an sgt distribution and that the error process $\epsilon_t= (\epsilon_{1t}, \ldots, \epsilon_{nt})'$ is a sequence of independent and identically distributed random vectors with zero mean and undit scale ($\sigma_i=1)$

The sgt distribution, introduced by \cite{Theodossiou1998}, is a general probability distribution that nests a large number of unimodal distributions, for example the normal distribution, as a special or limiting case. The sgt distribution can be applied to random variables that exhibit heavy tails or skewed errors, which is a common feature of financial and macroeconomic time series.

Under non-Gaussianity and independence of the error process $\epsilon_t$, it can be shown that matrix $B$ is unique apart from permutation and scaling of its columns. For a proof, see \cite{Lanne2017} (Appendix A). This means that there are $n!$ observationally equivalent matrices, and each permutation represents a reordering of the shocks, but which produce identical impulse response functions. From these matrices, we choose the permutation that maximizes the absolute value of the diagonal elements of $B$ and whose diagonal elements are positive, following the permutation convention of \cite{Pham1997}.

To ensure that the model is uniquely identified, we follow \cite{Anttonen2021} by restricting the parameter space such that no shock process follows a Gaussian distribution\footnote{Although no shock can follow a Gaussian distribution, a shock is allowed to be arbitrarily close to a Gaussian distribution. In the case that the shock process is close to Gaussian, the model may be only weakly identified.}.

For each structural shock $\epsilon_{it}$ ($i=1,\ldots, n$), the probability density function is defined as

\begin{equation}
f_i(\epsilon_{it}; \lambda_i, p_i, q_i) = \frac{p_i}{2 v_i q_i^{1/p_i}  \textit{Beta}\left(\frac{1}{p_i}, q_i \right) \left( \frac{\lvert \epsilon_{it}+m_i \rvert p_i}{q_i v_i^{p_i}\left( \lambda \text{sign}\left(e_{it}+m_i \right) + 1 \right)^{p_i}} +1 \right)^{\frac{1}{p_i}+q_i}},
\end{equation}

\noindent where $\lambda_i$ and $v_i$ are parameters that control skewness and scale, respectively, and $p_i$ and $q_i$ control the kurtosis of the distribution of the \textit{i}th shock. 

Moreover, we set 

\begin{equation}
m_i = \frac{2 v_i \lambda q_i^{1/p_i}  \textit{Beta}\left(\frac{2}{p_i}, q_i - \frac{1}{p_i} \right)}{\textit{Beta}\left(\frac{1}{p_i}, q_i \right)}
\end{equation}

\noindent such that each shock $\epsilon_{it}$ has a zero mean. Moreover, to ensure that the assumptions for identification holds, we set the parameter $v_i=1$, such that each component of the error process $\sigma_i$ has unit scale. In effect this allows for the possibility that a shock process may have non-finite variance. 

\subsection{Likelihood function and the prior distribution}

Following \cite{Lanne2017}, the likelihood function is expressed as

\begin{equation}
\label{ch3:eq:likelihood}
p \left(y \vert \theta 	\right) =   \lvert \det B \rvert ^{-T}  \prod_{i=1}^n \prod_{t=1}^T f_i \left( \iota_i' B^{-1} u_t \left( \Pi_t \right); \lambda_i, p_i, q_i \right),
\end{equation}

\noindent where $\theta_i = (\pi', \beta', \gamma')$, $\pi' = vec\left( \left[ a, A_1',\allowbreak \ldots, \allowbreak A_p' \right]' \right)' $, $\beta = vec(B)$ $\gamma = (\lambda_1,\allowbreak p_1,\allowbreak q_1,\allowbreak \ldots,\allowbreak \lambda_n,\allowbreak p_n,\allowbreak q_n)'$, $\iota_i$ is the \textit{i}th unit vector, and $u_t (\Pi) = y_t - a - A_1 y_{t-1} + \cdots +  A_p y_{t-p} $.

The posterior distribution $p(\theta \vert y )$ is proportional to the product of the likelihood function and the prior density $p(\theta)$, which may be formally expressed as

\begin{equation}
p (\theta \vert y) = \frac{p(y \vert \theta p(\theta)}{p(y)} \propto p(\theta) \lvert \det B \rvert ^{-T}  \prod_{i=1}^n \prod_{t=1}^T f_i \left( \iota_i' B^{-1} u_t \left( \Pi_t \right); \lambda_i, p_i, q_i \right),
\end{equation}

\noindent where $p(y)$ is the marginal likelihood.

The parameters $p_i$ and $q_i$ are assumed to have normal and log-normal prior distributions, respectively\footnote{In practice, the log transformations of $p_i$ and $q_i$ are considered to obtain more symmetric target distributions}. The prior mean of $q_i$ ($i=1,\ldots,n$) is set to 1, resulting in a relatively uninformative prior and low density near zero, with a non-infinite value implying a non-Gaussian distribution.

The prior for $p_i$ is set such that its posterior shrinks to near two, effectively shrinking the density function (\Cref{ch3:eq:likelihood}) to a skewed t distribution. The parameter $\lambda_i$ is assumed to have a uniform prior over the interval $[-1,1]$.

We use the tail exponent $\alpha_i = p_iq_i$ to describe the characteristics of the tails of the distributions for each shock $\epsilon_i$. The tail exponent has a similar interpretation as degrees of freedom in a t distribution, such that the distribution has a finite second moment if $\alpha_i>2$. In the empirical application, we use tail exponents to determine whether the assumption of non-finite variance is restrictive. 

We operate on the inverse of the impact matrix $B$, $vec(B^{-1}) \equiv b$ and assume that it follows a Gaussian prior distribution, $b \sim N(\underline{\matr{b}}, \underline{V}_{\matr{b}})$. We set $\underline{\matr{b}}=\matr{0}_{n^2}$ and an improper prior for $\underline{V}_{\matr{b}}$, such that $\underline{V}_{\matr{b}}=c_{\matr{b}} I_{n^2}$ and $c_{\matr{b}}^{-1}=0$. We assume a Minnesota-type prior for $\pi$, such that $\pi \sim N(\underline{\pi}, \underline{V}_{\pi})$. The prior variance for the \textit{ij}th element of $A_l$ ($l=1,\ldots, p$) is $v_{ij,l} = (\kappa_1/l^\kappa_3)^2$ for $i=j$ and $v_{ij,l} = (\kappa_1 \kappa_2 s_i/l^\kappa_3 s_j)^2$ for $i\neq j$, and the prior variance for the constant term $a$ is assumed to be $(s_i \kappa_4)^2$, where $s_i$ is the standard error of a $p$-lagged autoregression for the $i$th variable in $y$. In the estimation of the model, we consider an informative prior such that  $\kappa_1 = 0.2$, $\kappa_2 = \kappa_3 = 1$. The prior variance for the constant is assumed to be non-informative such that $\kappa_4 = 10,000$. We set the prior mean of the coefficient matrices such that that first coefficient for the first lag of each endogenous variable equal to one, i.e., $A_1$ is an identity matrix, and the prior means of other parameters are set to zero. 

The sample from the posterior distribution is obtained using the differential evolution Markov chain (DE-MC) algorithm of \cite{TerBraak2006} and \cite{TerBraak2008}, in which multiple chains are run in parallel. DE-MC is an adaptive MCMC algorithm that uses information from differences between past states of the parallel chains to update the chains. We refer to these papers for a detailed description of the algorithm.

In accordance with \cite{TerBraak2008} the convergence of the chains is assessed using the convergence statistic of \cite{Gelman2004}, according to which the chains are well converged when the the value of $\hat{R}$ is below the threshold value of 1.1 for each parameter. 	

\section{Macroeconomic effects of financial and uncertainty shocks}
\label{ch3:sec:macroeffects}

\subsection{Baseline model}
In this section, we present the results of our baseline model. The model is estimated using monthly data for the US covering the period 1976M7-2020M1\footnote{The data used in the estimation is described in detail in \cref{ch3:app:data}}. We use monthly data on manufacturing production, private employment, personal consumption expenditure, price index for personal consumption expenditure, stock market index. These variables enter the model in log differences. Moreover, we include the measure of macroeconomic uncertainty at three-month horizon of \cite{Jurado2015} (JLN), the excess bond premium (EBP) of \cite{Gilchrist2012} as measure of financial conditions and the 2-year US treasury yield.The uncertainty measure by \cite{Jurado2015} is the aggregate of the volatility of the forecast error of a large number of macroeconomic and financial time series. Although other observable measures or proxies of uncertainty exist, this measure is chosen due to its generality as it is not dependent on any single economic time series.

We estimate a SVAR(2) model\footnote{Choice of the lag length is based on standard information criteria. Estimating the model using more lags also hinders the convergence of the chains.} and define a Minnesota-type prior for the model parameters\footnote{The shrinkage parameter is set to 0.2}. The model is estimated based on 5,000,000 iterations with a thinning factor of 10 and a burn-in period of 400,000 iterations. Model inference is therefore based on a total of 100,000 observations. According to the \^{R} convergence statistic, the chain of iterations is well converged. 

\begin{figure}
\caption{Marginal posterior distributions of the tail exponents, baseline model}
     \centering
         \includegraphics[width=\textwidth]{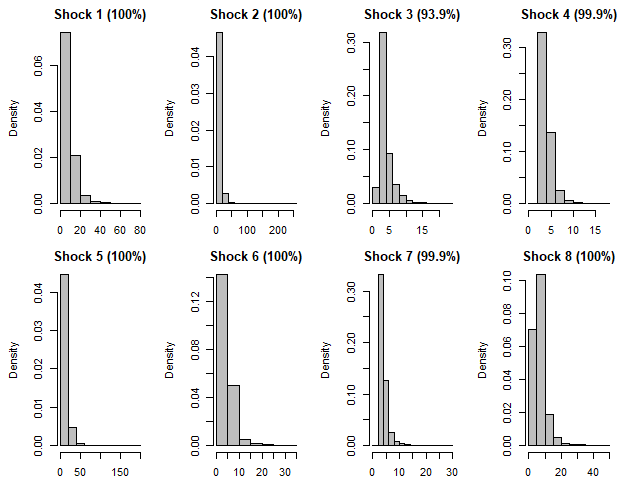}
         \label{ch3:fig:posterior_m}
         \begin{small}
         \caption*{\textit{Notes:} The tail exponent $\alpha_i = p_iq_i$ summarizes the characteristics of the tail of the posterior distribution. For each shock, the value in parenthesis is the probability that $\alpha>2$, or the probability that each shock has a finite variance. Generally, lower values imply that the distribution is fat-tailed, whereas $\alpha_i \rightarrow \infty$ implies a Gaussian distribution.}
		\end{small}
\end{figure}

To illustrate the estimated distributions for each shock, \cref{ch3:fig:posterior_m} presents marginal posterior distributions of their tail exponents. The figure shows that tail exponent for each shock is relatively low, that is the shocks are fat-tailed. This provides evidence in favor of the assumptions that the shock process is non-Gaussian and implies that the model is well identified. However, the probability that $\alpha_i>2$ for each shock is 94\%. This suggests that the structural shocks have a finite second moment with high probability, and therefore the assumption of non-finite variance is not restrictive for the estimated model.

Given that the model is statistically identified and no restrictions are imposed \textit{a priori}, the structural shocks have to be labelled based on extraneous information. Therefore, we label the financial and uncertainty shocks by examining the impulse response functions. \Cref{ch3:fig:irf_1_all} in \Cref{ch3:sec:appendix_figures} presents the impulse response functions to all structural shocks in the monthly specification.

The impulse response functions suggest that the first structural shock is the the uncertainty shock, given that the response of the uncertainty measure is non-zero on impact solely for this shock. Moreover, the second shock is the likeliest candidate to be the financial shock due to a non-zero initial response on the excess bond premium.

\begin{figure}
\caption{Impulse response functions to the uncertainty shock (structural shock 1), baseline model}
     \centering
         \includegraphics[width=\textwidth]{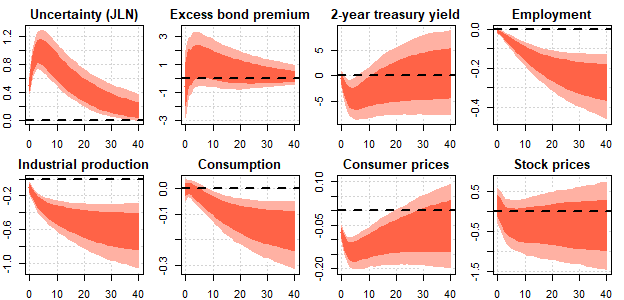}
         \label{ch3:fig:m1s1}
         \begin{small}
         \caption*{\textit{Notes:} The impulse response functions of the SVAR(2) model estimated using monthly data covering the period 1976M7--2020M1. The shaded areas denote the 68\% and the 95\% point-wise posterior intervals. The x-axis denotes months after the shock. The y-axis is in basis points for the uncertainty measure, the excess bond premium (EBP) and the 2-year treasury yield rate.  Macroeconomic variables and the stock market index are denoted in percentages. The scale of the shock is set to unity, corresponding to an estimate of a "typical" shock.}
\end{small}
\end{figure}

\begin{figure}
	\caption{Impulse response functions to the financial shock (structural shock 2), baseline model}         
    \centering
    \includegraphics[width=\textwidth]{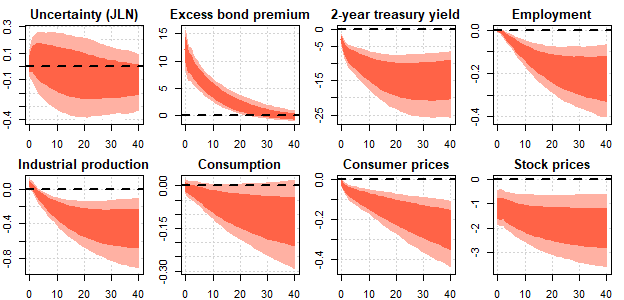}
	\label{ch3:fig:m1s2}
	\begin{small}
     \caption*{\textit{Notes:} See notes to \Cref{ch3:fig:m1s1}}.
	\end{small}        
\end{figure}

\Cref{ch3:fig:m1s1} and \Cref{ch3:fig:m1s2} present the impulse response functions and the 68\% and the 95\% point-wise posterior intervals for the uncertainty shock and the financial shock, respectively.

\Cref{ch3:fig:m1s1} shows the macroeconomic impact of a positive uncertainty shock and the 68\% and the 95\% point-wise posterior intervals. The shock has a hump-shaped response to the measure of uncertainty, whereby after the initial increase on impact, uncertainty continues to increase over a few periods, after which it gradually begins to diminish. The shock has a highly persistent effect on uncertainty, with a positive effect lasting even 40 months after the impact with high posterior probability. The increase in uncertainty leads to a persistent decline in economic activity by reducing employment, industrial production and consumption. It has a slightly positive effect on the excess bond premium, signifying a tightening of financial conditions. The shock also leads to a mild decrease in stock prices and the 2-year treasury yield.

\Cref{ch3:fig:m1s2} shows the macroeconomic impact of a contractionary financial shock and the point-wise posterior intervals. It has a persistent positive effect on the excess bond premium. However, with high posterior probability, its effect on uncertainty is non-zero in the short-term and slightly negative after 40 periods. The tightening of financial conditions also leads to a persistent decrease in employment, industrial production and consumption. Moreover, the shock also leads to a protracted decrease in consumer and stock prices and the 2-year treasury yield, although the effect is not non-zero with high probability.

The results suggest that the uncertainty shock may have a persistent negative effect on economic activity at least partly by tightening financial conditions, as evidenced by the increase in the excess bond premium and the decrease in stock prices. However, the financial shock does not appear to have a substantial effect on macroeconomic uncertainty. The results are then consistent with the macroeconomic literature that builds a link between uncertainty and financial conditions \citep{Arellano2018}.
These findings are also consistent with \cite{Caldara2016} and \cite{Furlanetto2019}, who report a similar macroeconomic impact of both the uncertainty and the financial shocks, based on the uncertainty measure by \cite{Jurado2015}. Remarkably, we do not find evidence that the financial shock meaningfully affects uncertainty, which is in line with the $\sigma$-EBP identification scheme, but not with the EBP-$\sigma$ scheme.

\subsection{Alternative specification using quarterly data}

In this section, we present the results of an alternative model estimated using quarterly data. The model is estimated using quarterly observations for the US covering the period 1976Q3-2020Q1. We include quarterly measures of gross domestic product (GDP), GDP deflator, employment, consumption, investment, stock market index. These variables enter the model in log differences. As in the baseline, we also include the quarterly average of the JLN uncertainty measure, the excess bond premium and the 2-year US treasury yield.

We estimate a SVAR(2) and use a Minnesota-type prior for the model parameters\footnote{As in the baseline model, the model is estimated based on 5,000,000 iterations with a thinning factor of 10 and a burn-in period of 400,000 iterations}. According to the \^{R} convergence statistic, the chain of iterations is well converged\footnote{For a few variables, the convergence statistic was slightly above the threshold value of 1.1, but it is not likely to affect the overall results.}. 

\begin{figure}
\caption{Marginal posterior distributions of the tail exponents, alternative model}
     \centering
         \includegraphics[width=\textwidth]{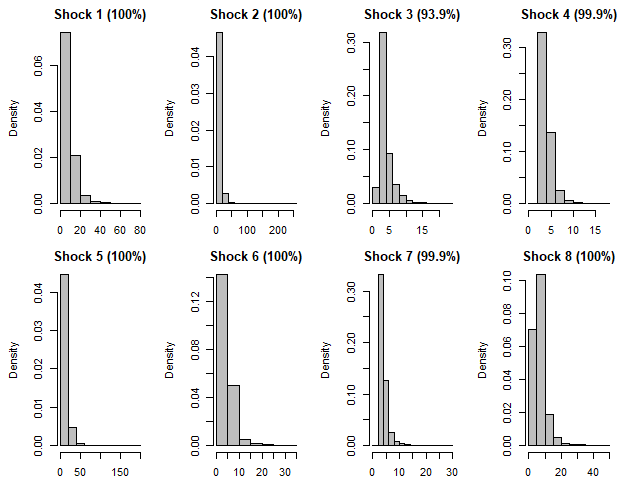}
         \label{ch3:fig:posterior_q}
         \begin{small}
         \caption*{\textit{Notes:} See notes to \Cref{ch3:fig:posterior_m}}.
         \end{small}
\end{figure}

The marginal posterior distribution of the tail exponents are illustrated in \Cref{ch3:fig:posterior_q}. For each shock the tail exponents imply that the shock process has fat tails and hence the model is well identified. The probability that $\alpha_i>2$ for each shock is 90\%, suggesting that finite error variances.

We label the shocks by examining the characteristics of the impulse response functions (see \Cref{ch3:fig:irf_q_all} in \Cref{ch3:sec:appendix_figures}). As in the baseline model, the impulse response functions suggest that the first structural shock is the the uncertainty shock and the second structural shock is the financial shock, given that the impact response on the uncertainty measure and excess bond premium are only non-zero for these shocks, respectively.

\begin{figure}
\caption{Impulse response functions to the uncertainty shock (structural shock 1), alternative model }
     \centering
         \includegraphics[width=\textwidth]{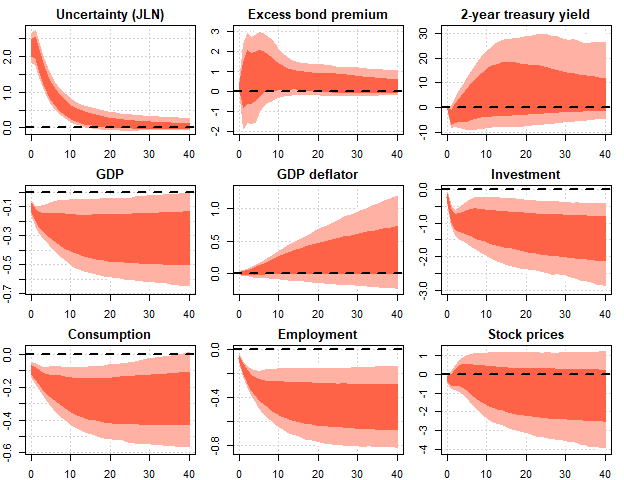}
         \label{ch3:fig:q1_m1s1}
         \begin{small}
         \caption*{\textit{Notes:} The impulse response functions of the SVAR(2) model estimated using quarterly data covering the period 1976Q3--2020Q1. The shaded areas denote the 68\% and the 95\% point-wise posterior intervals. The x-axis denotes quarters after the shock. The y-axis is in basis points for the uncertainty measure, the excess bond premium (EBP) and the 2-year treasury yield rate. Macroeconomic variables and the stock market index are denoted in percentages. The scale of the shock is set to unity, corresponding to an estimate of a "typical" shock.}
         \end{small}
\end{figure}

\begin{figure}
	\caption{Impulse response functions to the financial shock (structural shock 2), alternative model}         
    \centering
    \includegraphics[width=\textwidth]{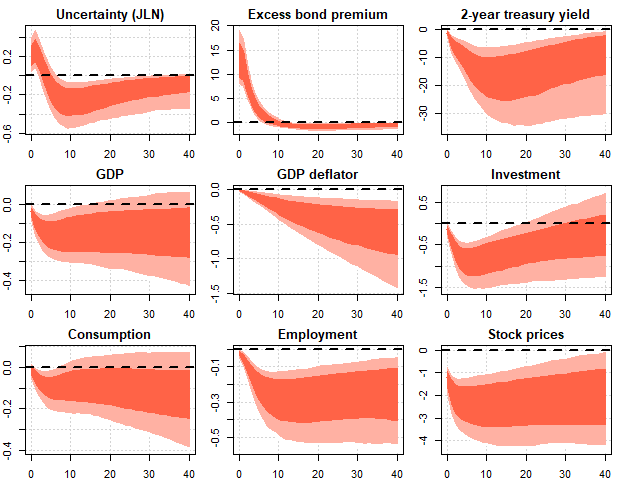}
	\label{ch3:fig:q1_m1s2}
	\begin{small}
     \caption*{\textit{Notes:} See notes to \Cref{ch3:fig:q1_m1s1}}
     \end{small}        
\end{figure}

The impulse response functions to the uncertainty shock and the financial shock are presented in \Cref{ch3:fig:q1_m1s1} and \Cref{ch3:fig:q1_m1s2}, respectively. \Cref{ch3:fig:q1_m1s1} shows the macroeconomic impact of a contractionary uncertainty shock and the 68\% and the 95\% point-wise posterior intervals. The shock has a highly persistent effect on uncertainty with high posterior probability. The effect on uncertainty is greatest on impact after which it begins to slowly taper out. The increase in uncertainty leads to a protracted decrease in GDP, investment, consumption and employment with high posterior probability. It also has a slightly positive but persistent effect on the excess bond premium and a mild decrease in stock prices, although the effect is not non-zero within the 68\% posterior interval in all horizons 

\Cref{ch3:fig:q1_m1s2} shows the macroeconomic impact of a contractionary financial shock and the point-wise posterior intervals. It has a persistent positive effect on the excess bond premium with high posterior probability, signifying the tightening of financial conditions. The shock also initially leads to a slight increase in uncertainty, but the effect dissipates quickly and turns negative within a few periods. Investment decreases initially, but then slowly begins to recover. The tightening of the financial conditions leads to a persistent decrease in employment, GDP and consumption. Moreover, the shock depresses stock prices and the 2-year treasury yield for an extended period with high posterior probability.

The impulse response functions are largely in line with the results obtained from the baseline model using monthly data, in that both the uncertainty and financial shocks have a persistent and independent negative effects on economic activity by reducing economic output, consumption and employment even after 40 quarters following the shock. Moreover, the results suggest that a contractionary financial shock leads to a tightening of financial conditions, whereas the financial shock does not affect macroeconomic uncertainty. These results suggest that the effects of contractionary uncertainty shock are more negative and persistent than those documented by \cite{Bonciani2019}. Moreover, in contrast with our findings,\cite{Furlanetto2019} find that the negative effect of the uncertainty shock on macroeconomic variables is short-lived, which could be partly explained  by their use of the VIX index as a measure of uncertainty, for which \cite{Caldara2016} find smaller macroeconomic effects compared to measures of macroeconomic uncertainty.
	
However, the two shocks have notably different effects on investment: The decrease in investment following a contractionary uncertainty shock seems to be greater and longer-lasting than after a contractionary financial shock. The same also applies to consumption, but to a lesser extent. This finding seems to suggest that the uncertainty shock affects investment and consumption through other channels besides affecting financial conditions. This supports the notion that an increase in uncertainty may affect firms' investment decisions through the value of "real options" (e.g., \cite{Bernanke1983}). Moreover, an increase in uncertainty may defer consumption by increasing households' precautionary savings.

\section{Conclusion} \label{ch3:sec:conclusion}

Uncertainty and financial shocks affect economic activity through multiple channels. We study the macroeconomic effect of financial and uncertainty shocks in the United States using a new approach by \cite{Anttonen2021}, where a the errors of the SVAR model follow a generalized t distribution. Due to non-Gaussianity of the error process, the structural shocks are statistically identified, and their economic interpretation is based on the properties of the impulse response functions.

The results confirm previous findings in the empirical literature that macroeconomic uncertainty and financial shocks are independent sources of business cycle fluctuations, but that uncertainty and financial conditions also interact dynamically. Namely, the uncertainty shock affects economic activity in part by tightening financial constraints, but a contractionary financial shock does not lead to an increase in uncertainty. Moreover, the results from the quarterly estimation suggest that the uncertainty shock may affect the economy beyond the business cycle by inducing a persistent negative effect on output and investment.

The significance of uncertainty and financial shocks in influencing business cycles and economic activity in the long-run makes understanding their macroeconomic effects highly relevant for academics and decision-makers. Studying the effects of the uncertainty shock on the long-run economic output provides an interesting avenue for future research. Specifically assessing the extent to which uncertainty affects productivity, production capacity and inputs of production is relevant to understanding its effects on potential economic output in the long-run as well as designing the appropriate policy measures to counteract the negative effects of uncertainty.

\cleardoublepage

\bibliographystyle{apalike}
\bibliography{ref3}  

%
%
%
%
\newpage

\section*{Appendices}

\begin{subappendices}

\begingroup
\setcounter{section}{0}    
\setcounter{subsection}{0}    

\def\thesubsection{\Alph{subsection}}

\subsection{Additional figures}
\label{ch3:sec:appendix_figures}

\setcounter{figure}{0}    
\renewcommand\thefigure{\thesubsection.\arabic{figure}}

 \begin{sidewaysfigure}
    \caption{Impulse response functions to structural shocks, baseline model} 
    \includegraphics[width=0.9\textwidth]{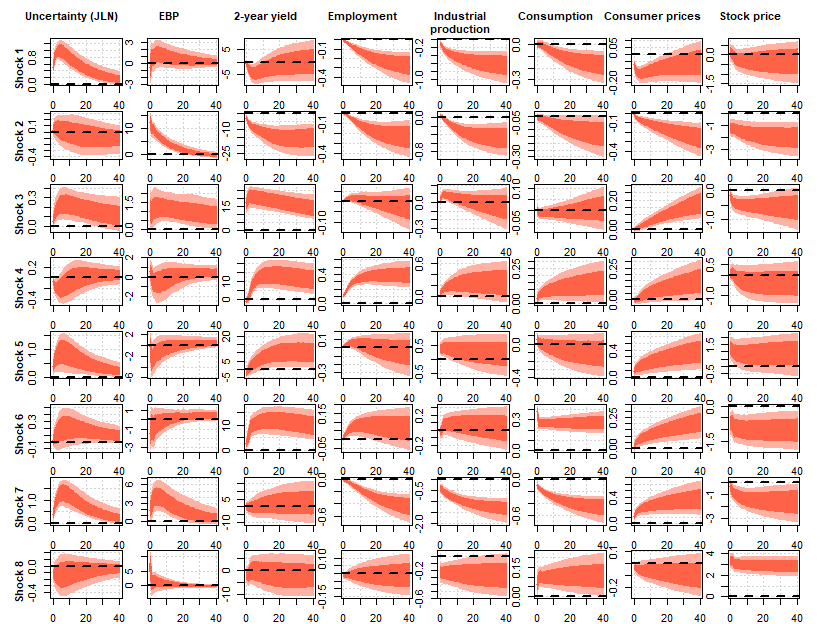}
	\caption*{\textit{Notes:} See notes to \Cref{ch3:fig:m1s1}}
	\label{ch3:fig:irf_1_all}    
\end{sidewaysfigure}

 \begin{sidewaysfigure}
   \caption{Impulse response functions to structural shocks, alternative model} 
   \includegraphics[width=0.9\textwidth]{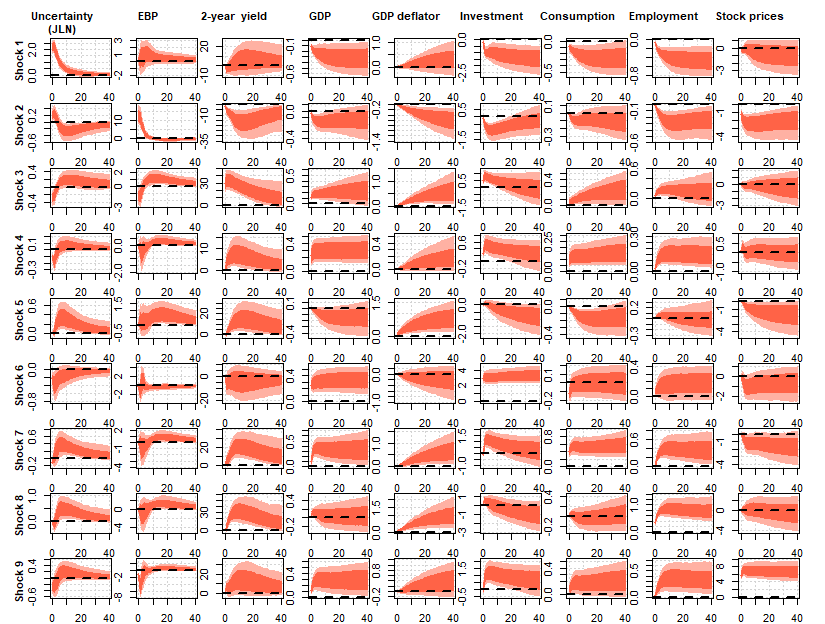}
\caption*{\textit{Notes:} See notes to \Cref{ch3:fig:q1_m1s1}}
   	   \label{ch3:fig:irf_q_all}   

\end{sidewaysfigure}

\subsection{Description of data}
\setcounter{table}{0}
\renewcommand{\thetable}{\thesubsection.\arabic{table}}

A detailed description of the data is provided in \Cref{ch3:tab:data}.

\label{ch3:app:data}
\begin{sidewaystable}[h]
\def\arraystretch{2}
  \centering
  \tiny
  \caption{Description of data}
    \begin{tabularx}{\linewidth}{p{2cm}p{6cm}XXXXp{4cm}}
    \toprule
    \textbf{Name} & \multicolumn{1}{l}{\textbf{Description}} & \textbf{Unit} & \textbf{Frequency} & \textbf{Adjustment} &  \textbf{FRED Code} & \textbf{Source} \\ \midrule 
        Consumption & Real personal consumption expenditures (chain-type quantity index) & Index (2012=100) & Monthly & Seasonally adjusted & DPCERA3M086SBEA & US. Bureau of Economic Analysis, FRED\\
    Consumption & Real Personal Consumption Expenditures & Billions of chained 2012 dollars & Quarterly & Seasonally adjusted annual rate & DPCERX & US. Bureau of Economic Analysis, FRED\\
    Consumption prices & Personal Consumption Expenditures: Chain-type Price Index & Index (2012=100) & Monthly & Seasonally adjusted & PCEPI & US. Bureau of Economic Analysis, FRED\\
    Employment & All Employees, Total Private & Thousands & Monthly & Seasonally adjusted & USPRIV & U.S. Bureau of Labor Statistics, FRED\\
    Employment & Nonfarm Business Sector: Employment for All Employed Persons & Index (2012=100) & Quarterly & Seasonally adjusted & PRS85006013 & U.S. Bureau of Labor Statistics, FRED\\
   Excess bond premium& Excess bond premium by \cite{Gilchrist2012} & Percent & Monthly & Not adjusted &  & \cite{Favara2016}, Board of Governors of the Federal Reserve System\\
    GDP & Real Gross Domestic Product &   Billions of Chained 2012 Dollars & Quarterly & Seasonally adjusted annual rate & GDPC1 & U.S. Bureau of Labor Statistics, FRED\\
    GDP deflator & Gross Domestic Product: Chain-type Price Index & Index (2012=100) & Quarterly & Seasonally adjusted & GDPCTPI & U.S. Bureau of Economic Analysis, FRED\\
    Industrial production & Industrial production: Manufacturing (NAICS) & Index (2017=100) & Monthly & Seasonally adjusted & IPMAN & Board of Governors of the Federal Reserve System (US), FRED\\
    Investment & Real Gross Private Domestic Investment& Billions of Chained 2012 Dollars  & Quarterly & Seasonally adjusted annual rate& GPDIC1 & U.S. Bureau of Economic Analysis, FRED\\
        Stock price & S\&P 500 index & Index & Month & Not adjusted & & S\&P\\
    Uncertainty & Measure for macroeconomic uncertainty (3-month ahead) by \cite{Jurado2015} & Index & Month & Not adjusted & & \url{sydneyludvigson.com}\\
    US 2-year yield & 2-Year Treasury Constant Maturity Rate & Percent & Month & Not adjusted & GS2 &Board of Governors of the Federal Reserve System (US), FRED\\ \bottomrule
\end{tabularx}%
  \label{ch3:tab:data}%
\end{sidewaystable}%
\endgroup

\end{subappendices}